\documentclass[12pt, a4paper]{article}

\usepackage[height=25cm,width=16cm]{geometry}
\usepackage{feynmp}
\usepackage{amsmath}
\usepackage{ascmac}
\usepackage{dcolumn}
\usepackage{bm,here}
\usepackage{subfig}
\usepackage{comment}
\usepackage{ifpdf}
\usepackage{slashed}
\usepackage{colortbl}
\usepackage{color}
\usepackage{ulem}
\usepackage{comment}
\usepackage{braket}
\usepackage[mathscr]{eucal}
\usepackage[sort&compress, numbers, merge]{natbib}
\usepackage{cancel}

\ifpdf
\usepackage{graphicx}     
\usepackage[bookmarksopen,colorlinks=true,linkcolor=bblue,citecolor=ppink,urlcolor=ppink]{hyperref}
\else     
\usepackage[dvipdfmx]{graphicx}     
\usepackage[dvipdfmx,bookmarksopen,colorlinks=true,linkcolor=bblue,citecolor=ppink,urlcolor=ppink]{hyperref}
\fi

\usepackage{multicol}
\definecolor{red}{rgb}{1,0,0}
\definecolor{ppink}{rgb}{0.921545,0.440586,0.687243}
\definecolor{bblue}{rgb}{0.400000,0.400000,1.000000}
\usepackage[charter]{mathdesign}
\usepackage{soul}
\usepackage{wrapfig}
\usepackage{arydshln}

\begin{document}

\begin{titlepage}

\begin{flushright}
\end{flushright}

\begin{center}

\vskip 1.5cm
{\large \bf
Indirect Detection of Dark Matter Around a Supermassive Black Hole \\
\vspace{0.07cm}
with High Energy-Resolution Gamma-Ray Telescopes
}


\vskip 2.0cm
{\large Yu Watanabe$^1$, Alexander Kusenko$^{1,2}$, and Shigeki Matsumoto$^1$}

\vskip 1.5cm
$^1${\sl Kavli IPMU (WPI), UTIAS, University of Tokyo, Kashiwa, 277-8583, Japan} \\ [.3em]
$^2${\sl Department of Physics and Astronomy, University of California, Los Angeles, California, 90095-1547, USA} \\ [.3em]

\vskip 2.5cm
\begin{abstract}
\noindent

We explore whether the unprecedented energy resolution of upcoming gamma-ray telescopes can uncover relativistic effects in photon spectra resulting from dark matter (DM) annihilation or decay near the supermassive black hole (SMBH) at the Galactic Center (GC), specifically, gravitational redshift, Doppler broadening due to Lorentz boosts, and kinetic energy enhancements arising from high DM velocities. By modeling DM density and velocity profiles under various SMBH formation scenarios and DM properties, we calculate the corresponding gamma-ray spectra and identify the conditions under which SMBH-induced spectral distortions become observable. We find that, in favorable cases, the observed spectra encode the DM velocity distribution near the SMBH, enabling potential discrimination among annihilation mechanisms with different velocity dependencies. Even when SMBH-induced effects are modest, the upcoming COSI mission, with sub-percent energy resolution surpassing the typical DM velocity dispersion at the GC, $\mathcal{O}(10^{-3})$, may still be able to detect subtle Doppler broadening. These results highlight a promising pathway for determining the origin of gamma-ray signals and probing DM properties through high-resolution spectral measurements.

\end{abstract}

\end{center}
		
\end{titlepage}

\tableofcontents
\newpage
\setcounter{page}{1}

\section{Introduction}
\label{sec: intro}

Revealing the nature of cosmic dark matter (DM) remains one of the most compelling unresolved problems in astrophysics, cosmology, and particle physics. Among various search strategies, indirect detection, observing standard model (SM) particles produced by DM annihilation or decay, is one of the most actively pursued approaches. A wide array of future experiments is planned to probe DM across a broad energy spectrum. In the MeV range, gamma-ray detection poses particular challenges due to dominant Compton scattering and complex astrophysical backgrounds. Currently, available data come only from COMPTEL\,\cite{schonfelder1993instrument} and SPI\,\cite{Winkler:2003nn}, whose sensitivities are poorer than those in adjacent energy bands, a limitation often referred to as the ``MeV gap.'' Recent technological and theoretical advances, however, have spurred several mission proposals aimed at bridging this gap, most notably NASA’s upcoming gamma-ray satellite COSI\,\cite{Tomsick:2021wed, Tomsick:2023aue}, scheduled for launch in 2027. At higher energies, the GeV range has been dominated by Fermi-LAT\,\cite{Fermi-LAT:2009ihh} for over a decade, with GAMMA-400\,\cite{Galper:2012fp} proposed as its successor. In the TeV domain, ground-based Cherenkov telescopes such as H.E.S.S.\,\cite{Bernlohr:2003tfz, Cornils:2003ve} and MAGIC\,\cite{Aleksic:2014poa, MAGIC:2014zas} have played central roles, and the forthcoming CTA\,\cite{Consortium:2010bc} is expected to significantly enhance their capabilities by offering improved sensitivity, broader energy coverage, and full-sky access through dual-hemisphere sites.

With the advent of advanced experiments, the prospects for discovering new signals, such as spectral lines at previously unknown energies, are becoming increasingly promising. However, even if such signals are observed, determining whether they originate from DM or from conventional astrophysical sources remains a major challenge. A key advantage of next-generation gamma-ray telescopes lies in their exceptional energy resolution. Instruments such as GAMMA-400 and CTA are expected to achieve energy resolutions at the percent level\,\cite{Galper:2013sfa, Maier:2019afm}, while COSI is anticipated to reach unprecedented sub-percent resolution\,\cite{Beechert:2022phz}. Importantly, signals induced by DM are often altered by the physical properties of DM itself, such as the Doppler broadening of spectral lines caused by the DM velocity distribution. If such broadening exceeds the intrinsic resolution of the detector, the effect becomes observable, and even a nominally monochromatic line may be distributed over multiple energy bins. This capability not only enhances the discrimination between DM and astrophysical origins but also allows one to extract information about the DM velocity profile or other characteristics from the spectral morphology. As we enter the era of high-resolution gamma-ray observations, exploring this avenue is both timely and crucial.

As a target for indirect detection, a DM-rich environment is highly desirable, most notably the Galactic Center (GC). A supermassive black hole (SMBH) resides at the location of Sgr~${\rm A^*}$\,\cite{Ghez:2008ms, Genzel:2010zy}, whose strong gravitational field leads to a significant accumulation of DM, thereby enhancing the potential signal\,\cite{Gondolo:1999ef, Regis:2008ij, Fields:2014pia}. Maintaining such high DM densities requires an increased velocity dispersion, which in turn amplifies velocity-suppressed annihilation processes, such as $p$-wave annihilation\,\cite{Shelton:2015aqa, Johnson:2019hsm}, forbidden annihilation\,\cite{Cannoni:2012rv, Cheng:2022esn}, and $s$-channel resonance-enhanced annihilation\,\cite{Arina:2014fna, Cheng:2023dau}. These enhancements make the SMBH environment a uniquely promising target for indirect searches. Beyond simply boosting the signal intensity, the SMBH also alters the spectral shape of the emitted photons through several relativistic effects: (i) gravitational redshift lowers the photon energy; (ii) large center-of-mass velocities induce Doppler broadening; and (iii) elevated DM relative velocities shift the photon energy upward. These effects are expected to imprint distinctive features on the observed spectrum, which may serve as valuable probes of DM properties.

In this article, we assess whether spectral modifications around the SMBH exceed the energy resolution of future telescopes, that is, whether DM near the SMBH produces distinctive photon fluxes with resolvable spectral features, and whether upcoming instruments can extract DM properties from them. The DM distribution around the SMBH depends on both its formation history and the nature of DM. We model the DM density and velocity profiles under several representative scenarios and compute the resulting gamma-ray spectra from DM annihilation or decay. We find that, in the case where the SMBH grows adiabatically within a cuspy halo of collisionless DM, next-generation instruments can probe Doppler broadening across the MeV to TeV energy range, assuming a thermal relic annihilation cross section of $\langle \sigma v \rangle = 10^{-26}\,{\rm cm^3/s}$. For smaller cross sections, even gravitational redshift and kinetic energy enhancements may become observable. On the other hand, in scenarios involving self-interacting DM with Coulomb-like forces, all spectral modifications are significantly amplified and potentially resolvable. Furthermore, the observed spectrum reflects the velocity distribution near the SMBH, enabling potential discrimination of annihilation mechanisms with different velocity dependence. Conversely, in scenarios where annihilation occurs with shallower spikes or where DM undergoes decay rather than annihilation, the outer halo dominates the observed flux, and SMBH-related effects become negligible. Nonetheless, the upcoming COSI mission, with its sub-percent energy resolution, which exceeds the typical DM velocity dispersion at the GC, $\mathcal{O}(10^{-3})$, may still be capable of detecting subtle Doppler broadening, even in the more conservative cases.

This article is organized as follows. In Section\,\ref{sec: Distribution}, we present the DM density and velocity distributions around the GC for various SMBH formation histories and DM scenarios. In Section\,\ref{sec: Signal}, we compute the resulting gamma-ray spectra and investigate whether SMBH-induced spectral features can be identified and utilized to constrain DM properties with future gamma-ray observatories. Finally, Section\,\ref{sec: summary} summarizes our findings.

\section{Impact of the SMBH on DM distribution}
\label{sec: Distribution}

In this section, we examine the distribution of dark matter (DM) density and velocity around the Galactic Center (GC), which depends on both the formation history of the supermassive black hole (SMBH) and the intrinsic properties of DM. We consider two representative scenarios: one with the collisionless DM and the other with the self-interacting DM.

\subsection{Collisionless DM scenario}
\label{subsec: Distribution Collisionless DM}

\begin{figure}[t]
    \centering
    \includegraphics[keepaspectratio, scale=0.4]{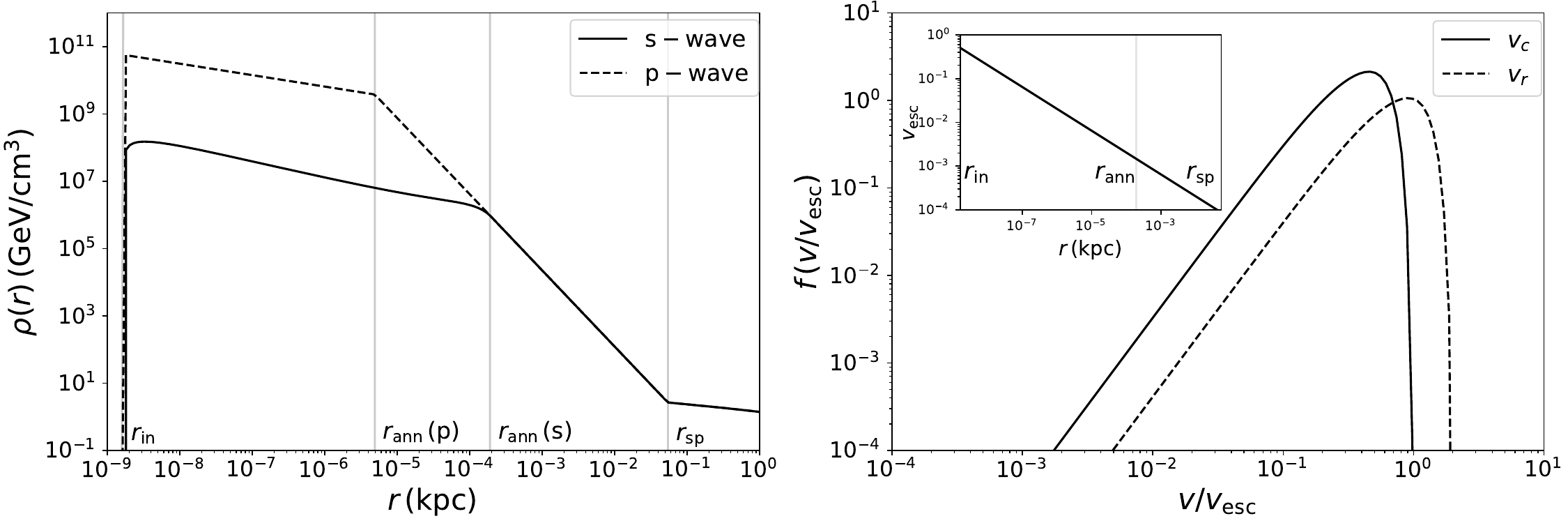}
    \caption{\small \sl {\bf Left:} DM density profile for the collisionless DM as a function of distance from the GC. {\bf Right:} Distributions of the relative and center-of-mass (c.o.m.) velocities, each normalized by the local escape velocity. The inset shows the escape velocity as a function of distance from the GC.
    }
    \label{fig: Distribution Collisionless DM}
\end{figure}

We assume that the SMBH grows adiabatically from a smaller seed embedded within a DM halo. We then briefly address alternative DM profiles that may arise from variations in DM properties and SMBH formation histories. The DM density profile can be divided into several regions, each exhibiting a distinct dependence on $r$, the distance from the GC:
\begin{equation}
    \rho_{\rm CDM} (r) =
    \left\{
    \begin{array}{lll}
        \rho_{\rm gNFW} (r) & \text{at} \quad r > r_{\rm sp} & \text{(Halo)}, \\
        \rho_{\rm sp} (r) & \text{at} \quad r_{\rm sp} \geq r > r_{\rm ann} & \text{(Spike)}, \\
        \rho_{\rm ann}(r) &  \text{at} \quad r_{\rm ann} \geq r > r_{\rm in} & \text{(Annihilation cusp)}, \\
        0 & \text{at} \quad r \leq r_{\rm in} & \text{(Capture)}.
    \end{array}
    \right.
    \label{eq: Distribution Collisionless DM}
\end{equation}
The behavior of these regions is illustrated in the left panel of Fig.\,\ref{fig: Distribution Collisionless DM}. In what follows, we provide a brief description of each region, proceeding from the outermost to the innermost.

\noindent
\pmb{Halo:}
In the absence of the SMBH, we assume that the DM halo is well described by the so-called generalized Navarro-Frenk-White (gNFW) profile\,\cite{Navarro:1996gj}, which is defined as
\begin{equation}
    \rho_{\rm gNFW} (r) = \rho_s \left( r/r_s \right)^{-\gamma} \left( 1+r/r_s \right)^{\gamma - 3}.
    \label{eq: gNFW}
\end{equation}
We adopt the following parameters\,\cite{Benito:2020lgu}: $\gamma = 0.2$, $\rho_s = 0.58\,\mathrm{GeV/cm^3}$, and $r_s = 40\,\mathrm{kpc}$.

\noindent
\pmb{Spike:}
Within the radius of gravitational influence of the SMBH, its strong gravitational field leads to a significant accumulation of dark matter (DM), resulting in a steeper density profile referred to as the spike profile\,\cite{Gondolo:1999ef}. According to Jeans' theorem, the steady-state solution of the collisionless Boltzmann equation can be expressed as a function of isolating integrals of motion, namely, the energy and angular momentum\,\cite{1962BAN....16..241O}. Consequently, the phase-space distribution function can be written as $f({\bf r}, {\bf v}) = f(E, L)$, where $E = \Psi(r) - |{\bf v}|^2/2$ is the relative energy per unit mass, and $L = | {\bf r}\times {\bf v}|$ is the magnitude of the angular momentum per unit mass. Here, ${\bf r}$ and ${\bf v}$ denote the position and velocity of DM particles, respectively, and $\Psi(r)$ is the positive-definite relative gravitational potential\,\cite{Lacroix:2018qqh}. The initial phase-space distribution function of DM associated with the gNFW profile in Eq.\,(\ref{eq: gNFW}) is given by\,\cite{Gondolo:1999ef}
\begin{equation}
    f(E,L) = \frac{\rho_s}{(2\pi \phi_0)^{3/2}} \frac{\Gamma (\beta)}{\Gamma (\beta-3/2)} \frac{\phi_0^\beta}{E^\beta},
\end{equation}
where $\beta = (6 - \gamma)/[2(2 - \gamma)]$ and $\phi_0 = 4 \pi G r_s^2 \rho_s / \left[(3 - \gamma)(2 - \gamma)\right]$, with $G$ denoting the Newtonian gravitational constant. The gravitational potential of the SMBH redistributes the dark matter density while conserving the radial action $I_r$ and the angular momentum, such that $I'_r(E', L') = I_r(E, L)$ and $L' = L$, where primed (unprimed) quantities refer to values after (before) the formation of the SMBH. For power-law profiles such as the adopted gNFW profile, the radial action is generally not obtainable in analytic form. Nevertheless, an approximate expression with an accuracy better than 8\% has been derived as follows\,\cite{Gondolo:1999ef}:
\begin{equation}
    I_r(E,L) =
    \frac{2\pi}{b}
    \left[
        -\frac{L}{\lambda}
        +\sqrt{2r_s^2 \phi_0} \left(\frac{E}{\phi_0} \right)^{\frac{4-\gamma}{2(2-\gamma)}}
    \right],
\end{equation}
where $\lambda = \left[2 / (4 - \gamma)\right]^{1 / (2 - \gamma)}$ and $b = \pi (2 - \gamma)/B(1 / (2 - \gamma), 3/2)$. On the other hand, when the gravitational potential is dominated by the SMBH, the radial action can be obtained analytically as $I'_r(E', L') = 2\pi \left(-L' + G M_{\rm BH}/\sqrt{2E'}\right)$, where $M_{\rm BH} \simeq 4.3 \times 10^6 M_\odot$ denotes the mass of the SMBH in our galaxy\,\cite{Gillessen:2017jxc}. By solving the conservation relations for $E$, we obtain the final phase-space distribution function as $f'(E', L') = f[E(E', L'), L]$. The corresponding density profile is then obtained by performing the integration over the phase space as
\begin{align}
    &
    \rho_{\rm sp} (r) =
    \int^{E'_m}_0 dE'
    \int ^{L_{\rm max}}_{L_{\rm min}} dL'\,
    \frac{4\pi L'}{r^2\,v_{\rm BH} }f'(E', L'),
    \quad
    v_{\rm BH} =
    \left[
        \frac{2GM_{\rm BH}}{r} - 2E' - \left(\frac{L'}{r}\right)^2
    \right]^{1/2},
    \nonumber \\
    &
    E_m' =
    \frac{G M_{\rm BH}}{r}
    \left( 1-\frac{4 R_s}{r}  \right),
    \quad
    L'_{\rm min} = 2 c R_s,
    \quad
    L'_{\rm max} =
    \left[
        2r^2 \left( \frac{G M_{\rm BH}}{r} - E'\right)
    \right]^{1/2},
    \label{eq: Integration}
\end{align}
where the parameters $R_s$ and $c$ denote the Schwarzschild radius and the speed of light, respectively. The integration can be performed semi-analytically, and the resulting density profile typically exhibits a power-law behavior with an index $\gamma_{\rm sp} = (9 - 2\gamma)/(4 - \gamma) \simeq 2.26$ in our case. The transition from the spike profile region to the outer halo profile, denoted by $r_{\rm sp}$, is defined by the condition $\rho_{\rm sp}(r_{\rm sp}) = \rho_{\rm gNFW}(r_{\rm sp})$, and it occurs at $r_{\rm sp} \simeq 54\,\text{pc}$.

\noindent
\pmb{Annihilation cusp:}
Consider DM with an annihilation cross-section $\langle \sigma v \rangle$, and define $r_{\rm ann}$ as the radius at which the condition $\rho_{\rm sp}(r_{\rm ann}) \langle \sigma v \rangle T_G/m_{\rm DM} = 1$ is satisfied, where $m_{\rm DM}$ denotes the DM mass and $T_G \sim 10^{10}\,{\rm yr}$ represents the age of the galaxy. In the region $r < r_{\rm ann}$, the annihilation timescale becomes shorter than the galactic age, leading to a suppression of the DM density\,\cite{Vasiliev:2007vh, Shapiro:2016ypb}. Under the assumption that DM particles on orbits entirely confined within $r_{\rm ann}$ eventually annihilate, the DM phase-space distribution is modified\,\cite{Shapiro:2016ypb}:
\begin{equation}
    f' = 
    \begin{cases}
        f'(E', L') & \text{at} \quad 0\leq E' \leq G M_{\rm BH}/r_{\rm ann}, \\
        0 & \text{at} \quad E' > G M_{\rm BH}/r_{\rm ann}.
    \end{cases}
\end{equation}
Integrating the modified phase-space distribution yields a density profile $\rho_{\rm ann}(r)$, which resembles that in the spike scenario but exhibits a weaker cusp, scaling as $\rho_{\rm ann} \propto r^{-1/2}$ in the case of $s$-wave annihilation. In contrast, $p$-wave annihilation requires a more detailed analysis based on the Boltzmann equation, which reveals an even shallower cusp, $\rho_{\rm ann} \propto r^{-0.34}$\,\cite{Shapiro:2016ypb}, and we adopt this result in our analysis. Owing to the velocity suppression inherent in the $p$-wave cross section, the corresponding $r_{\rm ann}$ is smaller than that in the $s$-wave case. As a representative example, we consider a dark matter mass of $m_{\rm DM} = 3\,{\rm MeV}$ and a canonical annihilation cross section of $\langle \sigma v \rangle = 10^{-26}\,{\rm cm^3\,s^{-1}}$ in the early Universe, with the temperature of the universe being $T \sim m_{\rm DM}/20$. Note that $\langle \sigma v \rangle$ refers to the thermally averaged total annihilation cross section, rather than that into any particular final state.

\noindent
\pmb{Capture:}
In the innermost region, $r < r_{\rm in} = 4R_s$, the density rapidly drops to zero due to capture by the SMBH. The above calculations have been performed within a non-relativistic framework. While some studies that incorporate relativistic effects suggest that the cusp may extend further inward to $r_{\rm in} = 2R_s$\,\cite{Sadeghian:2013laa}, as we will discuss in the next section, this difference does not affect our conclusions. We therefore adopt $r_{\rm in} = 4R_s$ in our analysis.

Next, we consider the distribution of the DM velocity. Suppose two DM particles, labeled 1 and 2, annihilate at a given position $r$. Their joint velocity distribution is described by the product $f'_1(r, {\bf v}_1)\, f'_2(r, {\bf v}_2)$, where $f'_i(r, {\bf v}_i)$ denotes the velocity distribution of particle $i$ at position $r$. It is useful to reformulate this distribution in terms of the relative velocity ${\bf v}_{\rm r} = {\bf v}_1 - {\bf v}_2$ and the center-of-mass velocity ${\bf v}_{\rm c} = ({\bf v}_1 + {\bf v}_2)/2$. Assuming that the gravitational potential is spherically symmetric, the one-dimensional distributions of $v_r = |{\bf v}_{\rm r}|$ and $v_c = |{\bf v}_{\rm c}|$ can be obtained by integrating over the appropriate angular variables as follows\,\cite{Lacroix:2018qqh, Yang:2024jtp}:
\begin{align}
    &
    f_{\rm ann}(r, v_r, v_c) \propto
    v_r^2\,v_c^2 \int^1_{-1}d(\cos\alpha) 
    \int^{2\pi}_0 d\phi
    \int^{\mu_0}_{-\mu_0} d(\cos\theta)\,
    f'_1(r, {\bf v_c} + {\bf v_r}/2)\,
    f'_2(r, {\bf v_c} - {\bf v_r}/2),
    \nonumber \\
    &
    \mu_0 = \frac{4 v_{\rm esc}^2 - 4v_c^2 - v_r^2}{4v_r v_c},
    ~~
    \cos \alpha = \frac{{\bf v}_c \cdot {\bf r}}{{\bf v}_c \times {\bf r}},
    ~~
    v_{\rm esc} = \sqrt{R_s/r},
    ~~
    |{\bf v}_{1,2}|^2 = v_c^2 + v_r^2/4 \mp v_c v_r \cos \theta .
\end{align}
Note that $f'_i(r, {\bf v}_1)$ is characterized not only by the energy $E_i$ but also by the angular momentum $L_i$ at $r \leq r_{\rm sp}$, and $L_i$ is expressed as functions of $v_r$, $v_c$, and the angular variables,
\begin{align}
    L_{1,2}^2 =
    r^2
    \left\{
        \frac{v_r^2}{4}\sin^2\theta \sin^2 \phi
        +
        \left[ \frac{v_r}{2} \cos \alpha \sin \theta \cos \phi \pm \sin \alpha \left( v_c \mp \frac{v_r}{2} \cos\theta \right) \right]^2
    \right\}.
\end{align}
Subsequently, the probability distributions of $v_r$ and $v_c$ are obtained by marginalizing the distribution $f_{\rm ann}$ over $v_c$ and $v_r$, respectively. Each distribution is then normalized to ensure that the total probability integrates to unity, which is presented in the right panel of Fig.\,\ref{fig: Distribution Collisionless DM}.

Finally, we consider alternative DM distributions that may arise due to variations in DM properties and the formation history of the SMBH. Although the initial DM distribution has been assumed to follow a gNFW profile, different parameter choices are possible. The resulting spike profile is only mildly sensitive to these variations. Specifically, for $0 < \gamma < 2$, the power-law index of the spike, $\gamma_{\rm sp}$, varies between approximately 2.25 and 2.50\,\cite{Gondolo:1999ef}, which has only a limited impact on our conclusions. Other DM profiles can also be considered. The Einasto profile, for instance, yields results that are similar to the gNFW case\,\cite{Shen:2023kkm}, whereas cored profiles, such as the isothermal profile, lead to shallower spikes characterized by $\gamma_{\rm sp} = 3/2$\,\cite{Gondolo:1999ef}. In addition, gravitational scattering of DM by a sufficiently dense and cuspy stellar population can heat the DM and soften the spike, resulting in an equilibrium profile with $\gamma_{\rm sp} = 3/2$\,\cite{Merritt:2003qk, Gnedin:2003rj, Shapiro:2022prq}. In the opposite extreme of instantaneous SMBH formation, referred to as the adiabatic growth limit, the resulting DM spike is significantly shallower, with $\gamma_{\rm sp} = 4/3$\,\cite{Ullio:2001fb}. Moreover, if the seed black hole is both sufficiently massive and initially offset from the center, the spike density can be further suppressed\,\cite{Ullio:2001fb}. It is worth noting here that, unlike the density profile, the velocity distribution in the spike region remains largely insensitive to such variations in DM properties and SMBH formation scenarios.

\subsection{Self-interacting DM scenario}
\label{subsec: Distribution Self-Interacting DM}

\begin{figure}[t]
    \centering
    \includegraphics[keepaspectratio, scale=0.4]{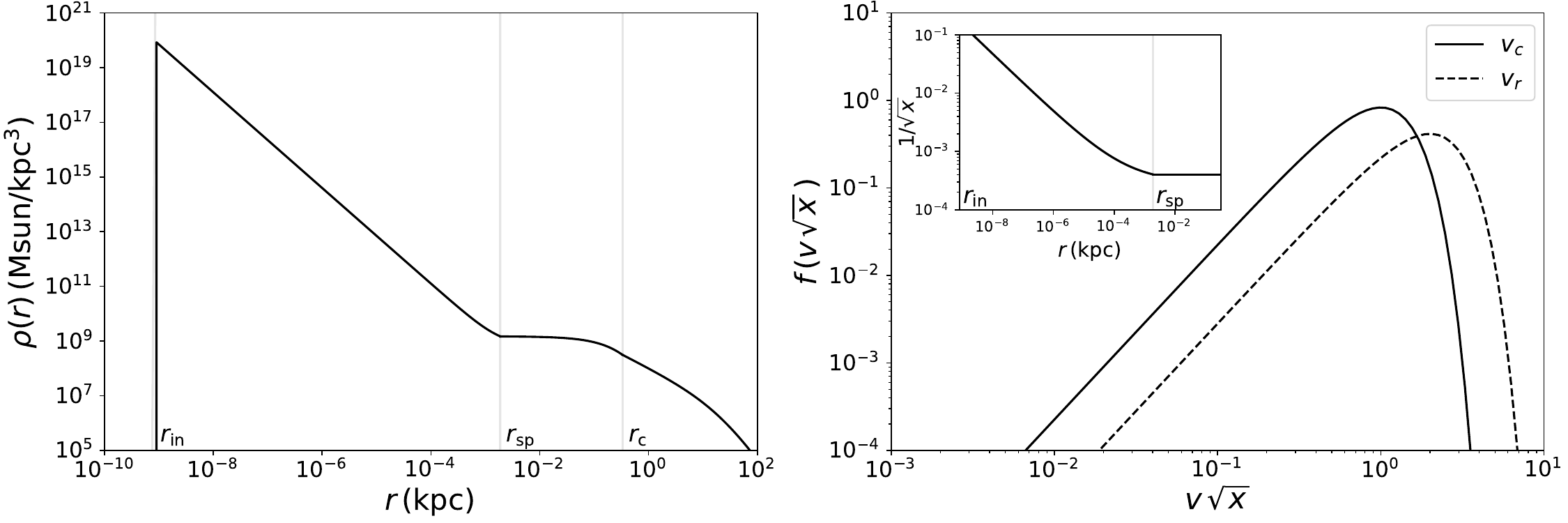}
    \caption{\small \sl DM distributions for the self-interacting DM scenario. {\bf Left:} DM density profile as a function of distance from the GC. {\bf Right:} Relative and c.o.m velocity distributions normalized by the typical velocity, which is shown in the inset as a function of distance from the GC.
    }
    \label{fig: Distribution Self-Interacting DM}
\end{figure}

Next, we consider the case of self-interacting DM. As in the previous case, the DM density profile can be divided into several regions, each exhibiting a distinct dependence on $r$:
\begin{equation}
    \rho_{\rm SIDM} (r) =
    \left\{
    \begin{array}{lll}
        \rho_{\rm NFW} (r) & \text{at} \quad r > r_{\rm c} & \text{(Halo)}, \\
        \rho_{\rm iso} (r) & \text{at} \quad r_{\rm c}
        \geq r > r_{\rm sp} & \text{(Core)}, \\
        \rho_{\rm sp}(r) &  \text{at} \quad r_{\rm sp} \geq r > r_{\rm in} & \text{(Spike)}, \\
        0 & \text{at} \quad r \leq r_{\rm in} & \text{(Capture)}.
    \end{array}
    \right.
    \label{eq: Distribution Collisionless DM}
\end{equation}
The behavior of these regions is illustrated in the left panel of Fig.\,\ref{fig: Distribution Self-Interacting DM}. In what follows, we provide a detailed description of each region, from the outermost to the innermost.

\noindent
\pmb{Halo:}
In the collisionless DM scenario, we assumed that the halo follows a gNFW profile, for which observational data typically favor a shallower slope with a small $\gamma$. One of the most appealing features of self-interacting DM is its potential to resolve the core–cusp problem by thermalizing in the GC\,\cite{Spergel:1999mh, Tulin:2017ara}. To investigate this possibility, we adopt a more cuspy profile at $r > r_{\rm c}$, specifically, the standard Navarro–Frenk–White (NFW) profile:
\begin{equation}
    \rho_{\rm gNFW} (r) =
    \rho_s
    \left( r/r_s \right)^{-1}
    \left( 1+r/r_s \right)^{-2},
\end{equation}
where the parameters $\rho_s$ and $r_s$ are $\rho_s = 4.1 \times 10^6 M_\odot\,{\rm kpc}^{-3}$ and $r_s = 26\,{\rm kpc}$\,\cite{Abazajian:2020tww, Alvarez:2020fyo}.

\noindent
\pmb{Core:}
Let us consider DM with a large self-interaction cross-section, denoted by $\langle \sigma_T v \rangle$, and define the critical radius $r_{\rm c}$ as the radius at which the condition $\rho_{\rm NFW}(r_{\rm c}) \langle \sigma_T v \rangle T_G/ m_{\rm DM} = 1$ is satisfied. In the region $r < r_{\rm c}$, DM typically undergoes multiple collisions over the timescale $T_G$, which leads to the formation of an isothermal DM core. The DM density profile in this region can be obtained by solving the Jeans equation\,\cite{Kaplinghat:2013xca, Kaplinghat:2015aga}, under the boundary conditions that the transition between the NFW profile and the isothermal core is continuous, and that the total enclosed mass within $r_{\rm c}$ remains the same as in the original NFW profile. As a representative example, we take the parameters $m_{\rm DM} = 3\,{\rm MeV}$ and $\langle \sigma_T v \rangle / m_{\rm DM} = 1.5\,{\rm cm^2\,g^{-1}\,km\,s^{-1}}$\,\cite{Kaplinghat:2015aga}. Under these assumptions, the resulting central density and velocity dispersion are $\rho_0 = 1.5 \times 10^9\,M_\odot\,{\rm kpc}^{-3}$ and $v_0 = 10^2\,{\rm km\,s^{-1}}$, respectively.

\noindent
\pmb{Spike:}
For $r < r_{\rm sp} = GM_{\rm BH}/v_0^2$, the gravitational force of the SMBH dominates, leading to the formation of a spike profile. Owing to self-interactions, the Boltzmann equations governing DM can be reduced to fluid conservation equations\,\cite{Shapiro:2014oha}. The density profile $\rho_{\rm sp}$ and the velocity dispersion $v_d$ are obtained by solving these equations under the boundary condition that the transition between the isothermal core profile and the spike profile remains continuous. The resulting solution typically exhibits a power-law behavior, given by $\rho_{\rm sp} \propto r^{-(3+a)/4}$ and $v_d \propto r^{-1/2}$, where $a$ denotes the velocity dependence index of the self-scattering cross-section, defined through $\sigma_T \propto v^{-a}$. In this work, we focus on the case of a Coulomb-like cross-section, corresponding to $a = 4$. Unlike the case of the collisionless DM scenario, the spike profile in the self-interacting scenario extends further inward, as DM is continuously transported from the outer regions due to its fluid-like behavior.

\noindent \pmb{Capture:}
In the innermost region, $r < r_{\rm in} = 2R_s$, the density rapidly drops to zero due to the DM capture by the SMBH\,\cite{Sadeghian:2013laa, Shapiro:2014oha}. Owing to the significant enhancement of the DM density by the spike profile, the region $r \sim r_{\rm in}$ plays a crucial role in indirect detection. Consequently, a fully relativistic calculation is required to model the DM behavior accurately.

Owing to the large cross section of the self-interaction among DMs and the significantly enhanced DM density profile near the SMBH, the DM fluid in this region is expected to be fully thermalized. Consequently, the DM velocity distribution follows the so-called Jüttner distribution\,\cite{1362825895258369152}. The joint probability distribution, expressed in terms of the relative velocity and the center-of-mass velocity, is given by the following formula\,\cite{Cannoni:2013bza, Cannoni:2015wba}:
\begin{equation}
    f_{\rm ann}(r, v_r, v_c) = \frac{x^2}{K_2^2(x)} \frac{\gamma_r^3 (\gamma_r^2 - 1)v_c^2}{(1-v_c^2)^2} e^{-x\sqrt{\frac{2+2\gamma_r}{1-vc^2}}},
\label{eq: Jutter}
\end{equation}
where $\gamma_r = 1/\sqrt{1 - v_r^2}$, $K_2(x)$ denotes the modified Bessel function of the second kind, and $x = m_{\rm DM}/T_{\rm DM}$, with $T_{\rm DM} = (3/2)\,m_{\rm DM} v_d^2$ representing the DM temperature. In the non-relativistic limit, Eq.\,(\ref{eq: Jutter}) reduces to the classical Maxwell--Boltzmann distribution, as illustrated in the right panel of Fig.~\ref{fig: Distribution Self-Interacting DM}. This approximation remains valid across most of the parameter space considered in this work, since the DM velocity dispersion $v_d$ does not exceed 0.1 even at the innermost radius $r_{\rm in}$. Nevertheless, for the sake of completeness, we employ the exact relativistic expression given in Eq.\,(\ref{eq: Jutter}) in the following section.

\section{Gamma rays from DM annihilation near the SMBH}
\label{sec: Signal}

In this section, we explore whether DM around the SMBH can produce distinctive photon fluxes with unique spectral features, and whether such features can be used to extract DM properties from future observations.
As a representative example, we consider the upcoming COSI experiment\,\cite{Tomsick:2021wed, Tomsick:2023aue}, a MeV gamma-ray observatory scheduled for launch in 2027.

\subsection{DM annihilation}
\label{subsec: Annihilating DM}

The mean photon flux produced by DM annihilation at radius $r$ is given as follows:
\begin{equation}
    \frac{d F_\gamma [r, E_\gamma]}{d E_\gamma} =
    \int dv_r\,dv_c\,
    f_{\rm ann}(r, v_r, v_c)\, (\sigma v_r)\,
    \Bigg[
        \sum_f\,{\rm Br}\,({\rm DM\,DM} \to f)\,
        \left.\frac{d N_\gamma}{dE_\gamma}\right|_f
    \Bigg],
    \label{eq: flux for 1 annihilation}
\end{equation}
where ${\rm Br}({\rm DM\,DM} \to f)$ is the branching ratio into a final state $f$, and $\left. dN_\gamma/dE_\gamma \right|_f$ denotes the photon spectrum produced in that channel. As a benchmark, we take $m_{\rm DM} = 3\,\mathrm{MeV}$ and $\langle \sigma v_r \rangle = 10^{-26}\,\mathrm{cm^3/s}$ in the early Universe, with the temperature $T \sim m_{\rm DM}/20$ (i.e., the so-called freeze-out temperature), where $\langle \sigma v_r \rangle$ denotes the thermal (velocity) average of the annihilation cross section. For simplicity, we consider only the $\gamma\gamma$ final state in the following discussion. The branching ratio ${\rm Br}({\rm DM\,DM} \to \gamma\gamma)$ sets the overall normalization of the flux, and we choose this value to lie within the sensitivity reach of COSI. As we will discuss later, variations in the final state or changes of variables do not significantly affect our conclusions. In the center-of-mass (c.o.m.) frame, DM annihilation into $\gamma\gamma$ yields a monochromatic photon line at $E_\gamma = \sqrt{s}/2$, where $\sqrt{s}$ is the total c.o.m. energy. Boosting to the lab frame with the c.o.m. velocity $v_c$ transforms this into a box-shaped spectrum:
\begin{equation}
    \left.\frac{d N_\gamma}{dE_\gamma}\right|_{\gamma \gamma} =
    \frac{2}{\sqrt{s}\gamma_c v_c} \left[
        \theta(E_{\gamma} - E_-)
        -\theta(E_{\gamma} - E_+)
    \right],
    \label{eq: fragmentation function}
\end{equation}
where $E_{\pm} = \sqrt{s}, \gamma_c (1 \pm v_c)/2$ are the upper and lower kinematic edges of the photon spectrum in the laboratory frame. Here, $\gamma_c = 1/\sqrt{1 - v_c^2}$ is the Lorentz factor, and $\theta(x)$ denotes the Heaviside step function, defined as $\theta(x) = 1$ for $x > 0$ and $\theta(x) = 0$ for $x < 0$. Then, the total flux from DM annihilation near the SMBH is obtained by integrating Eq.\,(\ref{eq: flux for 1 annihilation}) as
\begin{equation}
    \frac{d \Phi_\gamma (E_\gamma)}{d E_\gamma} \simeq
    \frac{1}{4\pi R^2}
    \frac{1}{2 f_{\rm DM} m_{\rm DM}^2}
    \int^{r_{\rm max}}_{r_{\rm in}} 4\pi r^2 dr\,
    \frac{\rho^2(r)}{g(r)}\,
    \frac{dF_\gamma [r, E_{\gamma}/g(r)]}{d E_\gamma}.
    \label{eq: flux produced}
\end{equation}
Here, $R \simeq 8.178$\,kpc denotes the distance from the Earth to the GC\,\cite{Gravity:2019nxk}, and $g(r) \equiv \sqrt{1 - R_s/r}$ accounts for the effect of gravitational redshift. The factor $f_{\rm DM}$ equals 1 for self-conjugate DM and 2 otherwise; in this article, we assume $f_{\rm DM} = 1$. The COSI detector has an angular resolution of approximately $5^\circ$\,\cite{Beechert:2022phz}, and we therefore consider observations within a cone of diameter smaller than this angle, corresponding to $r_{\rm max} \simeq 0.357\,\mathrm{kpc}$.

The finite energy resolution of the detector broadens the spectrum, resulting in the flux,
\begin{equation}
    \frac{d \Phi_{\rm obs} (E_{\gamma})}{d E_\gamma} =
    \int dE'_\gamma\,
    R_\epsilon(E_\gamma|E'_\gamma)\,
    \frac{d \Phi_\gamma (E'_\gamma)}{d E_\gamma}.
\end{equation}
where the kernel function $R_\epsilon(E_\gamma | E'_\gamma)$ is the so-called energy response function, representing the probability that a photon with true energy $E'_\gamma$ is reconstructed with an energy $E_\gamma$. This function can be approximated by a normalized Gaussian profile as follows\,\cite{Bringmann:2008kj}:
\begin{equation}
    R_\epsilon (E_\gamma|E'_\gamma) =
    \frac{1}{\sqrt{2\pi}\,\epsilon(E'_\gamma)\,E'_\gamma}\,\exp
    \left[-\frac{(E_\gamma - E'_\gamma)^2}{2[\epsilon(E'_\gamma)\,E'_\gamma]^2} \right],
    \label{eq: response function}
\end{equation}
with $\epsilon(E'_\gamma)$ denoting the fractional energy resolution. COSI is expected to achieve excellent line sensitivity, with a resolution (FWHM) given by $\epsilon(E\gamma) \approx 0.32\, (E_\gamma/\mathrm{MeV})^{-0.96}\,\%$\,\cite{Beechert:2022phz}. We identify three main mechanisms through which the SMBH affects the photon spectrum: (i) the deep gravitational potential induces redshift, reducing the photon energy by $\sim R_s/r$; (ii) the high center-of-mass velocity causes Doppler broadening by the Lorentz boost, increasing the spectral width by $\sim v_c$ and reducing the flux by $\sim 1/v_c \sim (r/R_s)^{1/2}$; (iii) the large relative velocity of DM enhances the photon energy by $\sim v_r^2 \sim R_s/r$, which shares the same radial dependence as the gravitational redshift but typically constitutes a subdominant effect. These effects become more pronounced for annihilations occurring deeper within the DM profile. In the following, we assess whether COSI is capable of probing these effects, i.e., whether the resulting spectral modifications exceed the instrument's energy resolution.

\subsubsection{Collisionless DM scenario}
\label{subsubsec: Collisionless DM}

\begin{figure}[t]
    \centering
    \includegraphics[keepaspectratio, scale=0.4]{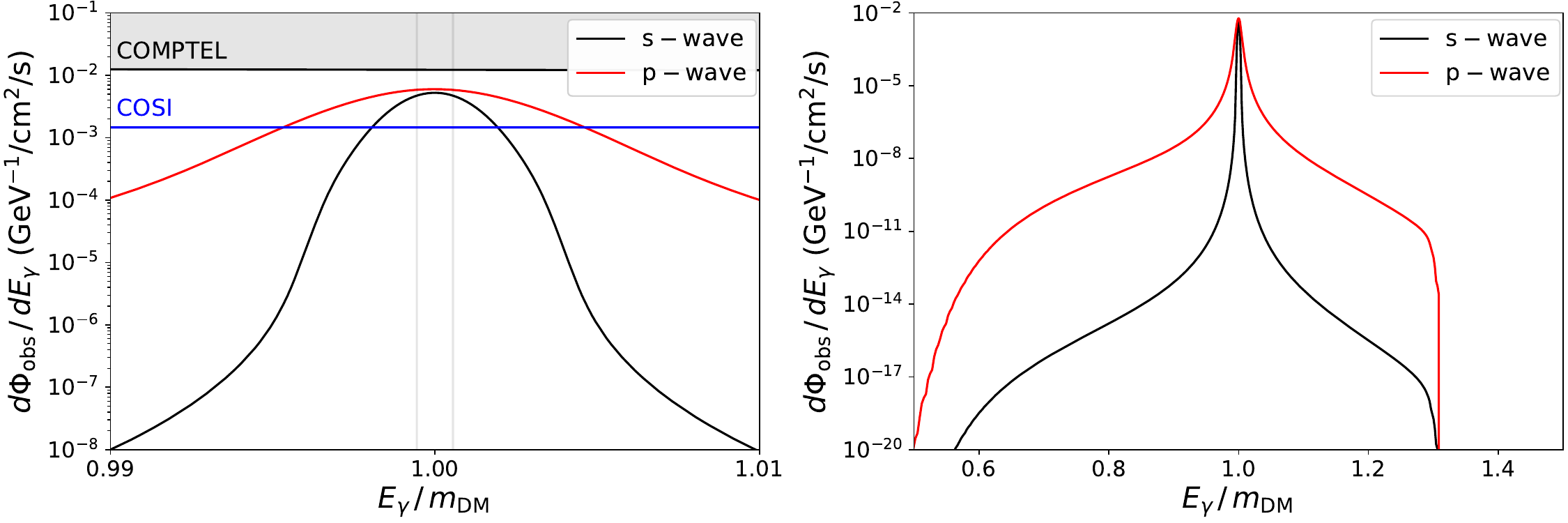}
    \caption{\small \sl {\bf Left:} Photon spectra from DM annihilation around the SMBH in the collisionless DM scenario, compared with the COMPTEL constraint and the projected sensitivity of COSI. The energy resolution of the COSI detector is indicated by the region enclosed between two light gray lines.
    {\bf Right:} The same spectra plotted over a wider energy range to illustrate broader spectral features.}
    \label{fig: flux Collisionless DM}
\end{figure}

We begin by considering the case in which the SMBH grows adiabatically from a smaller seed within a DM halo that follows a gNFW profile, as given in Eq.\,(\ref{eq: Distribution Collisionless DM}). The photon spectra resulting from DM annihilation, with properties specified above, are computed according to Eqs.\,(\ref{eq: flux for 1 annihilation})--(\ref{eq: response function}) and illustrated in Fig.\,\ref{fig: flux Collisionless DM}. The branching ratio ${\rm Br}({\rm DM\,DM} \to \gamma\gamma)$ is set to $1 \times 10^{-7}$ ($5 \times 10^{-6}$) for s-wave (p-wave) annihilation, serving as a reference value.

The quantitative features can be understood as follows. First, the DM density is significantly enhanced in the spike region, resulting in higher flux compared to the case without an SMBH. The flux is determined by the product of the J-factor and the annihilation cross section. The J-factor for an annular region at radius $r$ scales as $\rho_{\rm DM}^2(r)\,r^3$, and the annihilation cross section scales as $v_r^0$ for s-wave and $v_r^2$ for p-wave annihilation. The resulting flux is then Doppler-broadened by $v_c$. Consequently, including the effects of Doppler broadening and the velocity dependence of the annihilation cross section on $v_r$, the flux from an annular shell at radius $r$ scales as $\rho_{\rm DM}^2(r)\, r^{7/2}$ ($\rho_{\rm DM}^2(r)\, r^{5/2}$) in the s-wave (p-wave) case. Taking into account the DM density profile discussed in Sec.\,\ref{subsec: Distribution Collisionless DM}, the photon flux primarily originates from the region around $r_{\rm ann}$ in both s-wave and p-wave cases, which is found to be $2 \times 10^{-4}$ ($5 \times 10^{-6}$)\,kpc for s-wave (p-wave) annihilation. Meanwhile, the typical DM velocity is of order $\mathcal{O}(v_{\rm esc})$, as shown in Fig.\,\ref{fig: Distribution Collisionless DM}. At the radius $r_{\rm ann}$, we find $v_{\rm esc}(r_{\rm ann}) = 1.5 \times 10^{-3}$ ($9 \times 10^{-3}$) for the s-wave (p-wave) case, which exceeds the energy resolution of COSI. This implies that the Doppler broadening of the annihilation line can be resolved. Furthermore, since the typical DM velocities at $r_{\rm ann}$ differ between the s-wave and p-wave cases, COSI can, in principle, distinguish between the two annihilation modes by measuring the line width.

On the other hand, the gravitational redshift and the kinetic enhancement of photon energy both scale as $R_s/r$, whose contribution to the DM velocity is approximately $2.2 \times 10^{-6}$ in the s-wave case and $8.4 \times 10^{-5}$ in the p-wave case, respectively, at $r_{\rm ann}$. These values are smaller than the energy resolution of the COSI detector, implying that such effects cannot be directly resolved at these radii. Nevertheless, since general relativistic effects become increasingly significant at smaller radii, they still leave observable imprints on the gamma-ray spectrum, specifically, suppressed flux amplitudes, broadened line profiles, and extended low-energy tails due to the redshift. These spectral features are particularly pronounced in the p-wave scenario, owing to its smaller characteristic annihilation radius $r_{\rm ann}$. COSI improves upon the sensitivity of earlier MeV gamma-ray missions, such as COMPTEL\,\cite{schonfelder1993instrument, Schoenfelder:2000bu} and SPI\,\cite{Winkler:2003nn, Roques:2003xg}, by a factor of a few. As a result, COSI is expected to detect the Doppler broadening effect, though not gravitational redshift, due to the limited photon flux. Future MeV gamma-ray telescopes, including e-ASTROGAM\,\cite{e-ASTROGAM:2016bph}, AMEGO\,\cite{Kierans:2020otl}, and GECCO\,\cite{Orlando:2021get}, aim to improve the sensitivity by an additional order of magnitude. However, even these next-generation instruments may struggle to detect gravitational redshift effects, given the inherently low photon statistics from the innermost region near the SMBH.

\begin{figure}[t]
    \centering
    \includegraphics[keepaspectratio, scale=0.4]{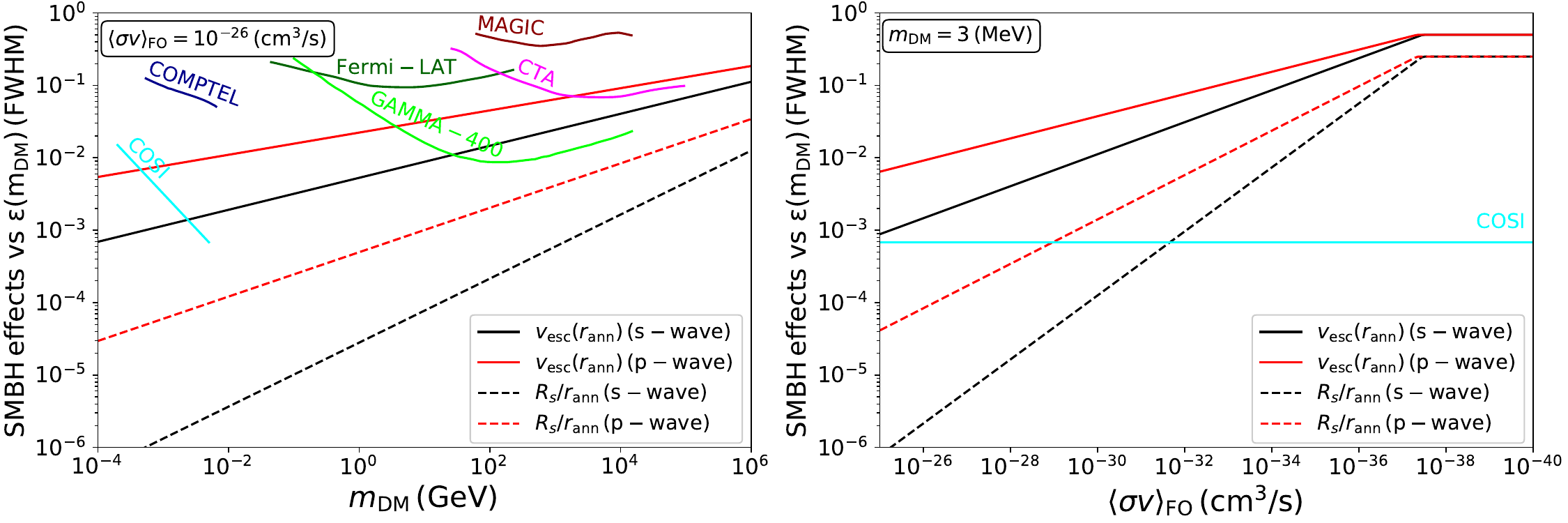}
    \caption{\small \sl Comparison of telescope energy resolutions with SMBH-induced spectral effects at $r_{\rm ann}$ (i.e., the typical radius of DM annihilation). Doppler broadening (gravitational redshift and kinetic enhancement) can be parametrized by $v_{\rm esc}$ ($R_s/r$). The annihilation cross section is fixed at $10^{-26}\,\mathrm{cm^3/s}$ (varied) while the DM mass is varied (fixed at $3\,\mathrm{MeV}$) in the left (right) panel.}
    \label{fig: change_variables}
\end{figure}

Next, we consider the impact of varying DM parameters. As a benchmark, we have so far adopted $m_{\rm DM} = 3\,\mathrm{MeV}$ and $\langle \sigma v \rangle = 10^{-26}\,\mathrm{cm^3/s}$ at the freeze-out epoch. We begin by examining the effect of changing $m_{\rm DM}$. As $m_{\rm DM}$ increases, the annihilation rate decreases, resulting in a smaller annihilation radius $r_{\rm ann}$, where the DM density is correspondingly higher. As discussed earlier, the photon flux predominantly originates from the vicinity of $r_{\rm ann}$, and the influence of the SMBH on the photon spectrum manifests through three primary mechanisms: Doppler broadening, gravitational redshift, and kinetic energy enhancement. The Doppler broadening is governed by the velocity, $v_c \sim v_{\rm esc}(r)$, while both gravitational and kinetic effects scale as $R_s/r$. We therefore evaluate whether these effects at $r_{\rm ann}$ exceed the energy resolution of upcoming $\gamma$-ray telescopes, considering both s-wave and p-wave annihilation cases. The results of this analysis are shown in the left panel of Fig.\,\ref{fig: change_variables}.

In the MeV range, the current telescope COMPTEL\,\cite{1992NASCP3137...85D} had a limited energy resolution of approximately 10\%, primarily due to its scintillator-based detection system. In contrast, upcoming instruments such as COSI are expected to achieve exceptional energy resolution below 1\%, owing to advanced Compton scattering techniques and the use of high-purity germanium detectors. In the GeV domain, GAMMA-400\,\cite{Galper:2012fp, Galper:2013sfa} improves upon the current Fermi-LAT telescope\,\cite{Fermi-LAT:2009ihh}, achieving an energy resolution of about 1\% by fully containing electromagnetic showers within a deep calorimeter. At TeV energies, telescopes such as MAGIC\,\cite{Sitarek:2015vba}, which rely on atmospheric Cherenkov imaging, are limited to energy resolutions of approximately 30\% due to the indirect nature of air-shower reconstruction. However, the upcoming CTA\,\cite{Consortium:2010bc, Maier:2019afm} is expected to significantly improve upon this, achieving resolutions of a few percent. As shown in Fig.\,\ref{fig: change_variables}, in the MeV and GeV energy ranges, Doppler broadening exceeds the instrumental resolution in both $s$-wave and $p$-wave scenarios. Hence, COSI and GAMMA-400 have the potential to detect SMBH-induced spectral effects and to distinguish between $s$-wave and $p$-wave WIMP annihilation.
In contrast, in the TeV regime, CTA is expected to probe the SMBH effect only in the $p$-wave case.

We also examine variations in the annihilation cross section, as illustrated in the right panel of Fig.\,\ref{fig: change_variables}. A decrease in $\langle \sigma v \rangle_{\rm FO}$ leads to a reduction in the annihilation radius $r_{\rm ann}$, thereby enhancing the detectability of SMBH-induced effects. When $r_{\rm ann}$ approaches the inner cutoff radius $r_{\rm in}$, the annihilation cusp disappears, resulting in a plateau characterized by $v_{\rm esc} = 0.5$ and $R_s/r = 0.25$. DM may also be produced via mechanisms other than freeze-out, such as freeze-in\,\cite{Hall:2009bx}, which typically predicts much smaller annihilation cross sections. In such scenarios, we find that COSI may be sensitive not only to Doppler broadening, but also to gravitational redshift and kinetic enhancement of photon energies.

Here, let us comment on more general velocity dependencies in DM annihilation. While we have so far focused on $s$-wave and $p$-wave annihilation as benchmark cases, DM annihilation may, in general, exhibit more intricate velocity structures. In fact, it is well known that light WIMPs at the MeV scale often require nontrivial velocity dependence\,\cite{Watanabe:2025pvc} to evade stringent constraints from cosmological observations\,\cite{Slatyer:2015jla, Kawasaki:2021etm}. Two well-studied alternatives are $s$-channel resonance\,\cite{Feng:2017drg, Bernreuther:2020koj, Binder:2022pmf, Brahma:2023psr} and forbidden channels\,\cite{DAgnolo:2015ujb, DAgnolo:2020mpt}. Our previous discussion can be naturally extended to these scenarios. In such cases, the characteristic radius $r_{\rm ann}$ typically shifts, and the spectral line width can be utilized to distinguish between different annihilation mechanisms. Even when $r_{\rm ann}$ remains unchanged, a stronger velocity dependence enhances the annihilation rate in the innermost region, thereby amplifying SMBH-induced spectral features. To be more precise, when the resonance or annihilation partner is sufficiently heavy in the $s$-channel resonant or forbidden-channel case, the annihilation process becomes kinematically accessible only in the vicinity of the SMBH, where DM particles attain large velocities. This provides a smoking-gun signature of the SMBH environment\,\cite{Cheng:2022esn, Cheng:2023dau}. Although our analysis has focused on the $\gamma\gamma$ final state, DM annihilation generally also yields continuum spectra, for instance through final-state radiation in $e^- e^+$ channels. The extension of our analysis to such final states is straightforward.

Finally, we consider alternative SMBH formation histories. As discussed in the previous section, the photon flux contribution from an annular shell located at radius $r$ scales as $\rho_{\rm DM}^2(r) \, r^{7/2}$ in the $s$-wave case and $\rho_{\rm DM}^2(r) \, r^{5/2}$ in the $p$-wave case, respectively. Thus, if the spike slope satisfies $\gamma_{\rm sp} \geq 7/4$ (s-wave) or $\gamma_{\rm sp} \geq 5/4$ (p-wave), the inner region provides the dominant contribution to the observable photon flux. On the other hand, in scenarios where the SMBH grows adiabatically from a smaller seed within a cuspy DM halo, such as a generalized NFW profile with a steeper inner slope or an Einasto profile, these conditions are typically satisfied, and the resulting spike enables the identification of SMBH-induced modifications to the DM distribution. Conversely, cored profiles, such as the isothermal profile, yield shallower spikes, making the corresponding gravitationally-induced spectral features more difficult to observe. As discussed in Sec.\,\ref{subsec: Distribution Collisionless DM}, other processes such as gravitational scattering by stars or instantaneous SMBH formation also predict shallower profiles, further diminishing the detectability of DM-induced spectral signatures. Nevertheless, even in such less favorable cases, the exceptional energy resolution of future telescopes like COSI may still permit the extraction of fine spectral features indicative of DM kinematics. COSI achieves $\mathcal{O}(10^{-4})$ energy resolution, as shown in Fig.\,\ref{fig: change_variables}, which is sufficient to resolve Doppler shifts induced by typical DM velocities at the GC of order $10^{-3}$. This implies that Doppler-induced spectral features may still be resolvable, allowing COSI to place constraints on the DM distribution parameters through precise measurements of line-shape distortions.

\subsubsection{Self-interacting DM scenario}
\label{subsubsec: Self-Interacting DM}

\begin{figure}[t]
    \centering
    \includegraphics[keepaspectratio, scale=0.4]{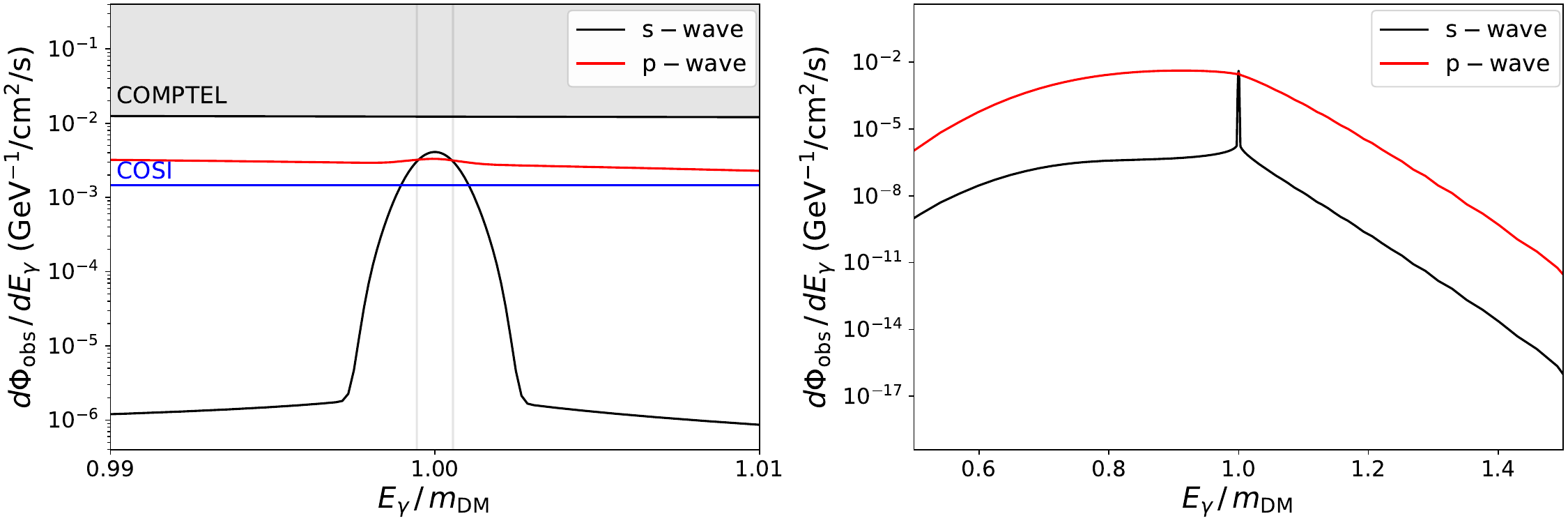}
    \caption{\small \sl {\bf Left:} Photon spectra from DM annihilation around the SMBH in the self-interacting DM scenario, compared with the COMPTEL constraint and the projected sensitivity of COSI. The energy resolution of the COSI detector is indicated by the region enclosed between two light gray lines.
    {\bf Right:} The same spectra plotted over a wider energy range to illustrate broader spectral features.}
    \label{fig: flux Self}
\end{figure}

We begin by considering the case where the self-scattering cross section scales as $\sigma_T \propto v^{-4}$, i.e., the form induced by a Coulomb-like force. The resulting photon spectra from DM annihilation, based on the parameters specified above, are shown in Fig.\,\ref{fig: flux Self}. The branching ratio ${\rm Br}({\rm DM,DM} \to \gamma\gamma)$ is set to $4 \times 10^{-7}$ for $s$-wave and $2 \times 10^{-2}$ for $p$-wave annihilation as benchmark values. In the spike region, the flux contribution from an annular shell at radius $r$ scales as $r^0$ for $s$-wave and $r^{-1}$ for $p$-wave annihilation, indicating a significant contribution from the innermost region. Unlike in the collisionless scenario, the spike in the self-interacting DM case extends all the way down to the inner cutoff radius $r_{\rm in}$ due to the fluid-like behavior induced by self-interactions. As a result, the SMBH has a more pronounced effect on the photon spectra than in the collisionless case: the spectra exhibit stronger Doppler broadening and a more prominent low-energy tail caused by gravitational redshift, both with enhanced flux amplitudes. However, since the density enhancement in the spike is relatively modest, the outer regions, namely the core and halo, dominate the total flux in the $s$-wave case, making it difficult for instruments with limited angular resolution, such as COSI, to probe SMBH-induced features. In contrast, for $p$-wave annihilation, the velocity dependence compensates for the lower density, leading to a dominant contribution from the inner region and a characteristic spectral signature shaped by the SMBH.

Next, we consider the impact of varying DM properties. As a benchmark, we have so far adopted $m_{\rm DM} = 3\,\mathrm{MeV}$ and $\langle \sigma_T v \rangle / m_{\rm DM} = 1.5\,\mathrm{cm^2/g} \times \mathrm{km/s}$. Increasing the DM mass or decreasing the self-scattering cross section reduces the core radius $r_{\rm c}$, leading to a higher DM density in the inner region and enhancing the flux contribution from that region. When $r_{\rm c}$ becomes smaller than the spike radius $r_{\rm sp}$, the DM density profile transitions to a hybrid configuration between the collisionless and self-interacting regimes: a collisionless spike forms at $r_{\rm sp}$, while deeper inside, within the spike or the annihilation cusp, a self-interacting spike with a slope $\gamma_{\rm sp} = -7/4$ develops. This intermediate regime inherits the advantages of both limits, namely a pronounced density enhancement and a small inner cutoff radius.

The above discussion can also be naturally extended to DM annihilation processes with alternative (non-trivial) velocity dependencies, such as $s$-channel resonance or forbidden channels, as well as to different final states, including final state radiation associated with $e^- e^+$ production. Furthermore, one may consider modified velocity dependencies for the DM self-scattering cross section. For instance, if the velocity dependence is weaker (e.g., scaling as $v^{-a}$ with $a < 4$), the resulting spike profile becomes shallower, and the flux is predominantly contributed by the outer region, thereby making it more difficult to isolate effects associated with the SMBH. Nevertheless, even in such scenarios, COSI may still offer sensitivity to DM properties, owing to its excellent energy resolution, particularly through the identification of Doppler-broadened spectral features originating from the halo and largely unaffected by SMBH-induced structures, as discussed in the previous section.

\subsection{Decaying DM}
\label{subsec: Decaying DM}

Next, we turn to the scenario in which DM undergoes decay rather than annihilation. The resulting photon flux can be evaluated using expressions analogous to Eqs.~(\ref{eq: flux for 1 annihilation})--(\ref{eq: response function}), with the essential difference that the flux is proportional to the DM density itself, rather than to its square. As a consequence, the enhancement of the signal due to the high DM density near the SMBH is less pronounced compared to the annihilation scenario.

The flux contribution from an annular shell at radius $r$ scales as $\rho_{\rm DM} \, r^{7/2}$. As discussed in Sec.\,\ref{subsec: Distribution Collisionless DM}, in realistic SMBH formation scenarios and DM models, the spike region is typically shallower than $r^{-7/2}$. As a result, the flux is dominated by contributions from the outer regions, which makes it more challenging to isolate effects specifically associated with the SMBH. Nevertheless, owing to its excellent energy resolution, COSI may still be capable of extracting information on DM properties, particularly through the identification of Doppler-broadened spectral features originating from the halo, as noted in the previous section.

\section{Summary}
\label{sec: summary}

Numerous upcoming gamma-ray telescopes are designed to probe DM across a broad range of energy scales, raising expectations for the detection of new signals. However, even if such signals are observed, distinguishing their DM origin from astrophysical backgrounds remains a major challenge. In this study, we explored the potential of a key advantage offered by next-generation instruments, their superior energy resolution, to tackle this issue, focusing on the region surrounding the SMBH at the GC. The presence of the SMBH can amplify DM-induced signals due to the high DM density and velocity in its vicinity. Furthermore, relativistic effects induced by the SMBH modify the photon spectrum in 3 ways: (i) gravitational redshift reduces the observed photon energy; (ii) large center-of-mass velocities lead to Doppler broadening via Lorentz boosts; and (iii) enhanced DM relative velocities increase the intrinsic photon energy. We investigated whether these spectral distortions exceed the energy resolution of future telescopes, i.e., whether DM near the SMBH generates photon fluxes with distinctive spectral features that can be disentangled from astrophysical backgrounds, enabling the extraction of DM properties by upcoming experiments.

The DM distribution around the SMBH depends on both its formation history and the nature of DM. We modeled the DM density and velocity profiles under several representative scenarios and computed the resulting gamma-ray spectra from DM annihilation or decay. In the case where the SMBH grows adiabatically within a cuspy halo of collisionless DM, the annihilation signal primarily originates from the region around $r_{\rm ann}$, where the annihilation cusp forms. Assuming a canonical annihilation cross section of $\langle\sigma v\rangle = 10^{-26}\,\mathrm{cm^3/s}$ during the freeze-out epoch, next-generation telescopes could detect Doppler broadening over a broad energy range from MeV to TeV, while gravitational redshift and kinetic energy effects would appear as flux suppression, potentially below detection thresholds. For smaller cross sections, $r_{\rm ann}$ shifts inward, and these relativistic effects may become directly observable.

On the other hand, in scenarios involving self-interacting DM with Coulomb-like forces, all spectral distortions are significantly enhanced due to the extended inner spike formed by fluid-like behavior. Furthermore, the observed spectrum reflects the velocity distribution near the SMBH, potentially enabling discrimination of the annihilation mechanism based on its velocity dependence. Conversely, in scenarios where annihilation occurs with shallower spikes or where DM undergoes decay instead of annihilation, the contribution from the outer halo dominates the observed flux, rendering SMBH-induced effects negligible. Nonetheless, the upcoming COSI mission, with its sub-percent energy resolution, exceeding the typical DM velocity dispersion at the GC, $\mathcal{O}(10^{-3})$, may still be capable of detecting subtle Doppler broadening originating from the halo and unaffected by SMBH-induced structures.


\section*{Acknowledgments}

All authors were supported by the World Premier International Research Center Initiative (WPI), MEXT, Japan (Kavli IPMU).
A.~K.\ was supported by the U.S.\ Department of Energy (DOE) under Grant No.\ DE-SC0009937, and by the Japan Society for the Promotion of Science (JSPS) KAKENHI under Grant No.\ JP20H05853.
S.~M.\ was supported by the Grant-in-Aid for Scientific Research from the Ministry of Education, Culture, Sports, Science and Technology, Japan (MEXT), under Grant No.\ 24H00244.
S.~M.\ and Y.~W.\ were supported by the Grant-in-Aid for Scientific Research from MEXT, under Grant No.\ 24H02244.

\bibliographystyle{unsrt}
\bibliography{refs}

\end{document}